\documentclass[journal,draftcls,onecolumn,12pt,twoside]{IEEEtranTCOM}
\usepackage{graphicx,cite,epsfig,amssymb,amsmath,subfigure,float}
\usepackage{graphicx,cite,epsfig}
\usepackage{amsmath,amssymb}
\usepackage{booktabs,multirow}
\usepackage{amsmath,bm}
\usepackage[noend]{algpseudocode}
\usepackage{algorithmicx,algorithm}
\usepackage{subfigure}
\usepackage{CJK}
\usepackage{color}
\usepackage{caption}
\usepackage{setspace}

\renewenvironment{cases}{\left\{\begin{array}[c]{ll}}{\end{array}\right.}

\begin{document}

\title{Modelling and Optimization of OAM-MIMO Communication Systems with Unaligned Antennas}

\author{Xusheng~Xiong,
        Hanqiong~Lou,
        Xiaohu~Ge,~\IEEEmembership{Senior~Member,~IEEE}



\thanks{\scriptsize{X.~Xiong, H.~Lou, X.~Ge (Corresponding author) are with the School of Electronic Information and Communications, Huazhong University of Science and Technology, Wuhan 430074, Hubei, P. R. China (e-mail: xiongxusheng@hust.edu.cn; louhq@hust.edu.cn; xhge@mail.hust.edu.cn).}}

}
\markboth{IEEE Transactions on Communications}%
{Submitted paper}

\maketitle




\begin{spacing}{1.34}
\begin{abstract}
The orbital angular momentum (OAM) wireless communication technique is emerging as one of potential techniques for the Sixth generation (6G) wireless communication system. The most advantage of OAM wireless communication technique is the natural orthogonality among different OAM states. However, one of the most disadvantages is the crosstalk among different OAM states which is widely caused by the atmospheric turbulence and the misalignment between the transmitting and receiving antennas. Considering the OAM-based multiple-input multiple-output (OAM-MIMO) transmission system with unaligned antennas, a new channel model is proposed for performance analysis. Moreover, a purity model of the OAM-MIMO transmission system with unaligned antennas is derived for the non-Kolmogorov turbulence. Furthermore, the error probability and capacity models are derived for OAM-MIMO transmission systems with unaligned antennas. To overcome the disadvantage caused by the unaligned antennas and non-Kolmogorov turbulence, a new optimization algorithm of OAM state interval is proposed to improve the capacity of the OAM-MIMO transmission system. Numerical results indicate that the capacity of OAM-MIMO transmission system is improved by the proposed optimization algorithm. Specifically, the capacity increment of the OAM-MIMO transmission system adopting the proposed optimization algorithm is up to 28.7\% and 320.3\% when the angle of deflection between the transmitting and receiving antennas is -24 dB and -5 dB, respectively.
\end{abstract}
\begin{IEEEkeywords}
Orbital angular momentum, multiple-input multiple-output, capacity, error probability, Laguerre-Gaussian.
\end{IEEEkeywords}
\end{spacing}
\IEEEpeerreviewmaketitle

\section{Introduction}
\label{sec1}
\IEEEPARstart {I}{n} recent years, Orbital Angular Momentum (OAM) technique \cite{1Allen92}, owing to its theoretically infinite OAM states and natural orthogonality among different OAM states, has attracted much attention in wireless communication systems. The OAM technique can provide a new degree of freedom for wireless communication systems, thereby increasing the system capacity of wireless communication systems \cite{2Ge17}. However, the orthogonality of OAM states is interfered by the atmospheric turbulence when the OAM technique is adopted in wireless communication systems \cite{3Xiong20}, \cite{4Lou19}. When the multiple-input multiple-output (MIMO) technology is used for OAM wireless communication systems, the misalignment between the transmitting and receiving antennas can cause the interference among different OAM states \cite{5Zhang13}. These two types of interference reduce the capacity of OAM-based MIMO (OAM-MIMO) transmission systems. Hence, it is a great challenge to improve the capacity of OAM-MIMO transmission systems considering the atmospheric turbulence and the misalignment between the transmitting and receiving antennas.

In 2004, the OAM technique was applied in optical wireless communications \cite{6Gibson04}. In optical wireless communications, the OAM beams were affected by the atmospheric turbulence, which led the energy leakage of the transmitted OAM states to the adjacent OAM states \cite{7Huang14}, \cite{8Sun16}. The crosstalk model of the Laguerre-Gaussian beams and the optimal OAM states set in atmospheric turbulence were investigated in \cite{9Anguita08}. Cheng {\em et al.} studied the non-diffraction Bessel-Gaussian beams in atmospheric turbulence and derived the conditional probability and the capacity of the non-diffraction Bessel-Gaussian beams \cite{10Cheng16}. The radial and normalized average power of vortex Gaussian beams have been theoretically formulated in consideration of weak to strong Kolmogorov atmospheric turbulence \cite{11Chen16}. Jiang {\em et al.} derived analytical formulas of the spiral spectrum of OAM beams in non-Kolmogorov turbulence \cite{12Jiang13}. OAM technique has also been applied in the radio frequency bands in recent studies. Thid\'{e} {\em et al.} employed the uniform circular array (UCA) antennas to first generate the OAM beams in the radio frequency bands \cite{13Thide07}. Tamburini {\em et al.} were the first to implement the experimental test using multi-mode OAM beams with states 0 and 1 in the microwave frequency bands, which indicated that the OAM technique could significantly improve the capacity of wireless communication systems \cite{14Tamburini12}. Zhang {\em et al.} utilized the partial phase plane reception to implement a 30.6 km long distance OAM transmission experiment \cite{15Zhang19}. Moreover, the millimeter-wave OAM beams were influenced by the atmospheric turbulence. The mode purity of millimeter-wave OAM beams is changed by the propagation distance and the value of OAM state \cite{16Cheng17}.

Another critical challenge in OAM beam transmission is that the transmitting and receiving antennas must be strictly aligned to ensure the maximum transmission rate. The misalignment between the transmitting and receiving antennas will lead to the distortion of the spiral wavefront. Furthermore, a part of the energy of OAM signals will be redistributed into the adjacent OAM states \cite{17Vas05}. Xie {\em et al.} utilized UCA antennas to generate OAM beams and verified that the distortion of OAM beams is increased with the increase of the angle of deflection \cite{18Xie16}. Chen {\em et al.} quantitatively investigated the effect of the misalignment between the transmitting and receiving antennas on OAM transmission systems and proposed a beam steering method to alleviate the degradation caused by unaligned antennas \cite{20Chen18}. The equivalent unitary matrix approach could be used to improve the spectrum efficiency of OAM transmission systems with unaligned antennas \cite{21Cheng19}. Chen {\em et al.} proposed the beam steering method with the estimated angle of arrival and amplitude to alleviate the degradation caused by unaligned antennas for OAM wireless communication systems \cite{22Chen20}.

In this paper, considering the effects of both atmospheric turbulence and misalignment between the transmitting and receiving antennas on the OAM-MIMO transmission system, the error probability and capacity of the OAM-MIMO transmission system are derived for performance analysis and an optimization algorithm is developed to improve the capacity. The contributions of this article are summarized as follows:

\begin{enumerate}
\item Considering the misalignment between the transmitting and receiving antennas, a channel model of the OAM-MIMO transmission system is established. Moreover, considering the non-Kolmogorov turbulence, a purity model of the OAM-MIMO transmission system with unaligned antennas is proposed. Based on the channel and purity models, the error probability and capacity models of the OAM-MIMO transmission system are derived.
\item Based on the capacity model of the OAM-MIMO transmission system with unaligned antennas in non-Kolmogorov turbulence, an OAM state interval optimization algorithm is designed to improve the capacity of the OAM-MIMO transmission system.
\item Numerical results indicate that the capacity of the OAM-MIMO transmission system adopting the optimization algorithm is improved. When the angle of deflection is -24 dB, the capacity increment of the OAM-MIMO transmission system adopting the optimization algorithm is up to 28.7\%. When the angle of deflection is -5 dB, the capacity increment of the OAM-MIMO transmission system adopting the optimization algorithm is up to 320.3\%.
\end{enumerate}

The rest of this article is organized as follows. In section II, the field distribution expression of the Laguerre-Gaussian (LG) beam with unaligned antennas is presented. Moreover, the channel model of the OAM-MIMO transmission system is proposed. In section III, the purity model of the OAM-MIMO transmission system with unaligned antennas in non-Kolmogorov turbulence is proposed. In section IV, the error probability and capacity models of the OAM-MIMO transmission system with unaligned antennas in non-Kolmogorov turbulence are derived. Furthermore, the optimal OAM state interval is developed to improve the capacity of the OAM-MIMO transmission system. In section V, the numerical results are analyzed and discussed. In the end, conclusions are drawn in Section VI.

\section{System Model}
\label{sec2}

\subsection{Channel model with aligned antennas}

When the traveling-wave antennas are used to generate the OAM beams, the strength of OAM beams can be described by LG beams \cite{23Yao13}. In cylindrical coordinate systems, the LG beam is expressed as \cite{1Allen92}
\[u_{LG}^{_{p,l}}(r,\phi ,z) = a\sqrt {\frac{{p!}}{{\pi (p + \left| l \right|)!}}} \frac{1}{{{\omega _l}(z)}}{\left( {\frac{{\sqrt 2 r}}{{{\omega _l}(z)}}} \right)^{\left| l \right|}}{e^{ - {{\left( {\frac{r}{{{w_l}(z)}}} \right)}^2}}}L_p^{\left| l \right|}\left( {\frac{{2{r^2}}}{{\omega _l^{\rm{2}}(z)}}} \right){e^{i(\left| l \right| + 2p + 1)\zeta (z)}}{e^{\frac{{ - i\pi {r^2}}}{{\lambda {R_l}(z)}}}}{e^{ - il\phi }},\tag{1a}\]
with
\[{\omega _l}\left( z \right) = {\omega _l}\sqrt {1 + {{\left( {\frac{z}{{{z_R}}}} \right)}^2}},\tag{1b}\]
\[{R_l}(z) = z\left[ {1 + {{\left( {\frac{{\pi \omega _l^2}}{{\lambda z}}} \right)}^2}} \right],\tag{1c}\]
where $r$ denotes the radial distance, $\phi$ denotes the azimuthal angle, $z$ denotes the propagation distance, $l$ denotes the value of OAM state, $a$ denotes the complex constant and $i$ denotes the imaginary unit. $p$ denotes the radial index, $p$ is configured as $p = 0$ for the proposed OAM-MIMO communication systems in this paper. ${\omega _l}\left( z \right)$ denotes the beam waist radius with the OAM state $l$ and propagation distance $z$. When $z = 0$, ${\omega _l}\left( z \right)$ can be denoted as ${\omega _l}$. ${z_R} = \frac{{\pi \omega _l^2}}{\lambda }$ denotes the Rayleigh distance, $\lambda $ denotes the wavelength. $L_p^{\left| l \right|}\left( {\frac{{2{r^2}}}{{\omega _l^{\rm{2}}(z)}}} \right)$ denotes the generalized Laguerre polynomial and $L_p^{\left| l \right|}\left( {\frac{{2{r^2}}}{{\omega _l^{\rm{2}}(z)}}} \right) = 1$ when $p = 0$. $\zeta (z) = \arctan \left( {\frac{z}{{{z_R}}}} \right)$ denotes the Gouy phase. ${R_l}(z)$ denotes the curvature radius of the OAM spiral wavefront. ${e^{ - il\phi }}$ denotes the helical phase distribution. The radius of the circle region with the maximum energy strength is expressed as \cite{25Wang17}
\[{r_{\max }}(z) = \sqrt {\frac{{\left| l \right|}}{2}} {\omega _l}\left( z \right) = {\omega _l}\sqrt {\frac{l}{2}\left( {1 + {{\left( {\frac{z}{{{z_R}}}} \right)}^2}} \right)}. \tag{2}\]

\begin{figure}
  \centering
  \includegraphics[width=14cm]{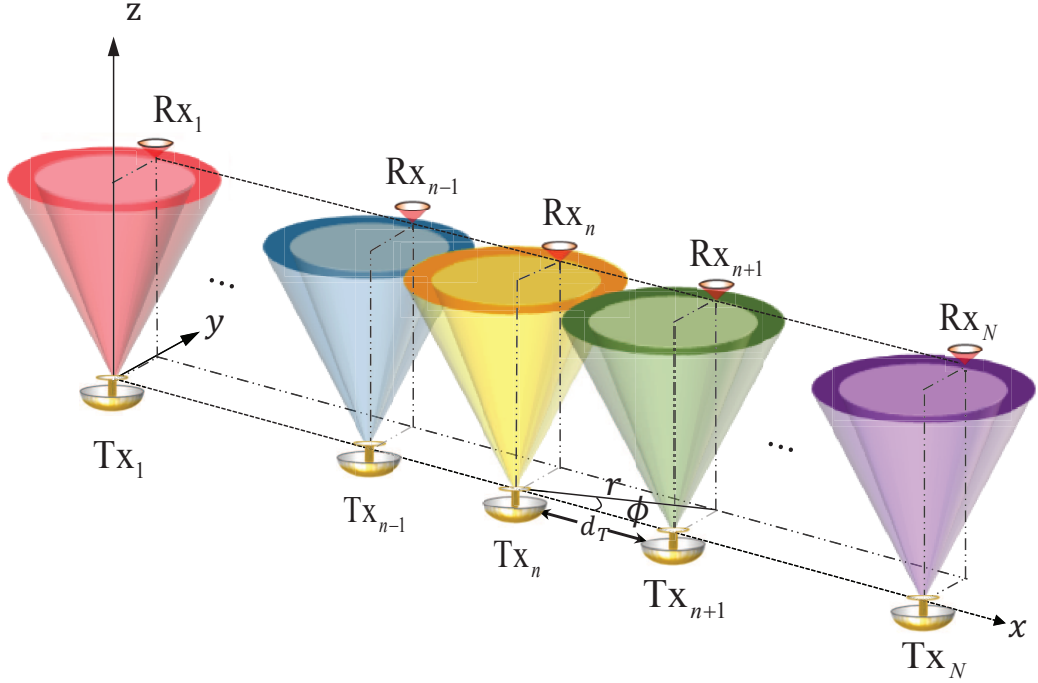}
  \caption{System model.}\label{Fig. 1}
\end{figure}

Assuming that there are $N$ transmitting antennas in OAM-MIMO communication systems. The $N$ transmitting antennas are placed as a uniform linear array. Every transmitting antenna can generate $L$ OAM states simultaneously and the $L$ OAM states are equidistant. As shown in Fig. 1, the cylindrical coordinate system is used to express the position of the antennas, where ${d_T}$ stands for the distance between the adjacent transmitting antennas. Similarly, the distance between the adjacent receiving antennas is also ${d_T}$. Assuming that the position of $T{x_1}$ is the origin of the cylindrical coordinate system as $(0,0,0)$. For the $n - th$ transmitting antenna $T{x_n}$, the position can be expressed as $((n - 1){d_T},0,0)$. Assuming that the receiving antennas lie in the plane of $z = {d_{TR}}$, so the position of the first receiving antenna $R{x_1}$ can be expressed as $({r_{\max }}(z),\frac{\pi }{2},{d_{TR}})$.

When the transmitting and receiving antennas are perfect aligned, the azimuth angle between the $i - th$ transmitting antenna $T{x_i}$ and the $j - th$ receiving antenna $R{x_j}$ is expressed as
\[{\phi _{ji}} = \begin{cases}
\arctan \frac{{{r_{\max }}\left( z \right)}}{{\left| {i - j} \right|d_T }}, & j > i \\
\frac{\pi }{2}, & j = i \\
\pi  - \arctan \frac{{{r_{\max }}\left( z \right)}}{{\left| {i - j} \right|d_T }}, & j < i \\
\end{cases}. \tag{3}\]

The radial distance in the cylindrical coordinate system between the $i - th$ transmitting antenna $T{x_i}$ and the $j - th$ receiving antenna $R{x_j}$ is denoted as
\[{r_{ji}} = \sqrt {{{\left( {\left| {i - j} \right|{d_T}} \right)}^2} + r_{\max }^2(z)}. \tag{4}\]
Then the distance between the $i - th$ transmitting antenna $T{x_i}$ and the $j - th$ receiving antenna $R{x_j}$ is expressed as
\[{d_{ji}} = \sqrt {d_{TR}^2 + {{\left( {\left| {i - j} \right|{d_T}} \right)}^2} + r_{\max }^2(z)}. \tag{5}\]

The channel response between the $i - th$ transmitting antenna $T{x_i}$ and the $i - th$ receiving antenna $R{x_i}$ is as follows
\[h_{LG,ii}^l = \beta \frac{\lambda }{{4\pi {d_{ii}}}}{e^{ - ik{d_{ii}}}}{e^{ - i\frac{\pi }{2}l}}. \tag{6}\]
The channel response between the $i - th$ transmitting antenna $T{x_i}$ and the $j - th$ receiving antenna $R{x_j}$ is as follows
\[h_{LG,ji}^l = {\beta _{LG,ji}}\frac{\lambda }{{4\pi {d_{ji}}}}{e^{ - ik{d_{ji}}}}{e^{ - i{\phi _{ji}}l}}, \tag{7}\]
where $\lambda $ denotes the wavelength, $k = \frac{{2\pi }}{\lambda }$ is the wavenumber.

In this paper, the transmitting antennas equipped with traveling-wave ring resonators are used to generate OAM beams with different OAM states. The traveling-wave antennas are assumed to be equipped with carefully designed reflectors \cite{26Hui15}, \cite{27Zhang16}. Even though the transmitting antennas are equipped with reflectors, the divergence angles of OAM signals with different OAM states are slightly different \cite{28Zheng15}. Nevertheless, the differences in the divergence angles will make the proposed OAM-MIMO transmission system very complicated for theory analysis. Because of the reflectors used in the transmitting antennas and the feasibility of the analysis, the size of the circle region is assumed to be equal in the following theoretical analysis \cite{2Ge17}, which is expressed as
\[{\omega _l}\sqrt {\frac{{\left| l \right|}}{2}\left( {1 + {{\left( {\frac{z}{{{z_R}}}} \right)}^2}} \right)}  = {\omega _{l'}}\sqrt {\frac{{\left| {l'} \right|}}{2}\left( {1 + {{\left( {\frac{z}{{{z_R}}}} \right)}^2}} \right)}. \tag{8}\]
Substituting the Rayleigh distance ${z_R} = \frac{{\pi \omega _l^2}}{\lambda }$ into (8), (8) can be expressed as
\[A\omega _{l'}^4 - B\omega _{l'}^2 + C = 0, \tag{9a}\]
where
\[A = \omega _l^2\left| l \right|{\pi ^2}, \tag{9b}\]
\[B = \left| l \right|\left( {{\pi ^2}\omega _l^4 + {z^2}{\lambda ^2}} \right), \tag{9c}\]
\[C = \omega _l^2\left| {l'} \right|{z^2}{\lambda ^2}. \tag{9d}\]

The beam waist ${\omega _{l'}}$ of OAM signal with OAM state $l'$ can be obtained with the solution of (9a). Substituting ${\omega _{l'}}$ into (1a), the strength distribution of OAM signal with OAM state $l'$ can be derived. If $u_{LG}^{p,l}\left( {r,\phi ,z} \right)$ stands for the response of OAM electromagnetic wave in the cylindrical coordinate system after an input of a unit pulse, the response at the $i - th$ receiving antenna $R{x_i}$ can be expressed as $u_{LG}^{ii} = h_{LG,ii}^l\hat x$ when the OAM signal is transmitted by the $i - th$ transmitting antenna $T{x_i}$. When the OAM signal is transmitted by the $i - th$ transmitting antenna $T{x_i}$, the response at the $j - th$ receiving antenna $R{x_j}$ can be expressed as $u_{LG}^{ji} = h_{LG,ji}^l\hat x$, where $\hat x$ stands for the unit pulse input. Since $\hat x$ is all the same, the following equation can be obtained as
\[\frac{{u_{LG}^{ii}}}{{u_{LG}^{ji}}} = \frac{{h_{LG,ii}^l}}{{h_{LG,ji}^l}}. \tag{10}\]
Then the ${\beta _{LG,ji}}$ is expressed as
\[{\beta _{LG,ji}} = \beta \frac{{{d_{ji}}}}{{{d_{ii}}}}{\left( {\frac{{{r_{ji}}}}{{{r_{\max }}\left( z \right)}}} \right)^{\left| l \right|}}{e^{ - \frac{{r_{ji}^2 - r_{\max }^2\left( z \right)}}{{\omega _l^2\left( z \right)}}}}{e^{ - i\frac{{\pi \left( {r_{ji}^2 - r_{\max }^2\left( z \right)} \right)}}{{\lambda {R_l}\left( z \right)}}}}{e^{ik\left( {{d_{ji}} - {d_{ii}}} \right)}}. \tag{11}\]
The channel response between the $i - th$ transmitting antenna $T{x_i}$ and the $j - th$ receiving antenna $R{x_j}$ is derived as
\[h_{LG,ji}^l = \beta \frac{\lambda }{{4\pi {d_{ii}}}}{\left( {\frac{{{r_{ji}}}}{{{r_{\max }}\left( z \right)}}} \right)^{\left| l \right|}}{e^{ - \frac{{r_{ji}^2 - r_{\max }^2\left( z \right)}}{{\omega _l^2\left( z \right)}}}}{e^{ - i\frac{{\pi \left( {r_{ji}^2 - r_{\max }^2\left( z \right)} \right)}}{{\lambda {R_l}\left( z \right)}}}}{e^{ - ik{d_{ii}}}}{e^{ - i{\phi _{ji}}l}}. \tag{12}\]

\subsection{Channel model with unaligned antennas}

It is difficult to fully align the transmitting and receiving antennas in practical OAM-MIMO transmission systems. When the transmitting and receiving antennas are misaligned, how to improve the capacity of OAM-MIMO transmission systems poses new challenges. As shown in Fig. 2, there are three types of misalignments between the transmitting and receiving antennas: a) lateral displacement between the transmitting and receiving antennas; b) angular deflection between the transmitting and receiving antennas; c) combination of lateral displacement and angular deflection between the transmitting and receiving antennas.
\begin{figure}
  \centering
  \includegraphics[width=16cm]{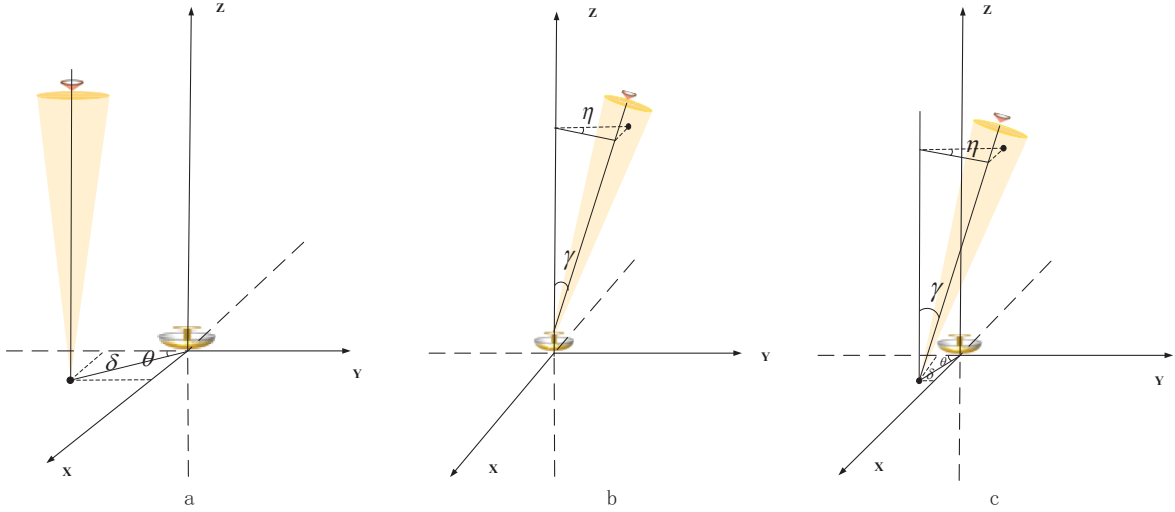}
  \caption{(a)lateral displacement, (b)angular deflection, (c)combination of lateral displacement and angular deflection.}\label{Fig. 2}
\end{figure}

When the lateral displacement happens between the transmitting and receiving antennas, the LG beam can be expressed as \cite{17Vas05}
\[u_D^{_{p,l}}(r,\phi ,z) = \frac{{u_{LG}^{_{p,l}}(r,\phi ,z)}}{{{\omega _l}(z)}}{\left( {r{e^{i\phi }} - \delta {e^{i\theta }}} \right)^l}\exp \left( { - \frac{{{r^2} + {\delta ^2}}}{{\omega _l^2\left( z \right)}}} \right)\sum\limits_{m =  - \infty }^\infty  {{I_m}\left( {\frac{{2r\delta }}{{\omega _l^2\left( z \right)}}} \right)} \exp \left[ {im\left( {\phi  - \theta } \right)} \right], \tag{13}\]
where $\delta $ denotes the radial displacement, $\theta $ denotes the azimuth angle, ${I_m}( \cdot )$ denotes the modified Bessel function of the first kind of integer order $m$.

When the angular deflection happens between the transmitting and receiving antennas, the LG beam can be expressed as \cite{17Vas05}
\[u_A^{_{p,l}}(r,\phi ,z) = \frac{{u_{LG}^{_{p,l}}(r,\phi ,z)}}{{{\omega _l}(z)}}{\left( {r{e^{i\phi }}} \right)^l}\exp \left( { - \frac{{{r^2}}}{{\omega _l^2\left( z \right)}}} \right)\sum\limits_{m =  - \infty }^\infty  {{J_m}\left( {\varsigma r} \right)} \exp \left[ {im\left( {\phi  - \eta  + \frac{\pi }{2}} \right)} \right], \tag{14}\]
where $\gamma $ denotes the angle of deflection, $\eta $ denoted the azimuth angle, ${J_m}( \cdot )$ denotes the Bessel function of the first kind of integer order $m$, $\varsigma $ is related to $\gamma $.

When the combination of lateral displacement and angular deflection happens between the transmitting and receiving antennas, the LG beam can be expressed as \cite{17Vas05}
\[\begin{aligned}
u_{DA}^{_{p,l}}(r,\phi ,z) = &\frac{{u_{LG}^{_{p,l}}(r,\phi ,z)}}{{{\omega _l}(z)}}{\left( {r{e^{i\phi }} - \delta {e^{i\theta }}} \right)^l}\exp \left( { - \frac{{{r^2} + {\delta ^2}}}{{\omega _l^2\left( z \right)}}} \right)\sum\limits_{m =  - \infty }^\infty  {{I_m}\left( {\frac{{2r\delta }}{{\omega _l^2\left( z \right)}}} \right)} \\
&  \times \sum\limits_{n =  - \infty }^\infty  {{J_n}} \left( {\varsigma r} \right)\exp \left[ {im\left( {\phi  - \theta } \right) + in\left( {\phi  - \eta  + \frac{\pi }{2}} \right)} \right]
\end{aligned}. \tag{15}\]

Assuming that only the receiving antennas are moved and the movement happens in the plane $z = {d_{TR}}$. The azimuth angle between the $i - th$ transmitting antenna $T{x_i}$ and the $j - th$ receiving antenna $R{x_j}$ is expressed as
\[\phi _{ji}^{DA} = \begin{cases}
\arctan \frac{{\delta \sin \theta  + {d_{TR}}\tan \gamma \sin \eta  + {r_{\max }}(z)}}{{\delta \cos \theta  + {d_{TR}}\tan \gamma \cos \eta  + \left| {i - j} \right|{d_T}}}, & j > i  \\
\arctan \frac{{\delta \sin \theta  + {d_{TR}}\tan \gamma \sin \eta  + {r_{\max }}(z)}}{{\delta \cos \theta  + {d_{TR}}\tan \gamma \cos \eta }}, & j = i  \\
\pi  - \arctan \frac{{\delta \sin \theta  + {d_{TR}}\tan \gamma \sin \eta  + {r_{\max }}(z)}}{{\delta \cos \theta  + {d_{TR}}\tan \gamma \cos \eta  + \left| {i - j} \right|{d_T}}}, & j < i  \\
\end{cases}. \tag{16}\]
When the combination of lateral displacement and angular deflection happens between the transmitting and receiving antennas, the radial distance in the cylindrical coordinate system is denoted as
\[{r_{DA,ji}} = \sqrt {{{\left( {\left| {i - j} \right|{d_T} + {d_{TR}}\tan \gamma \cos \eta  + \delta \cos \theta } \right)}^2} + {{\left( {{r_{\max }}(z) + {d_{TR}}\tan \gamma \sin \eta  + \delta \sin \theta } \right)}^2}}. \tag{17}\]
Then the distance between the $i - th$ transmitting antenna $T{x_i}$ and the $j - th$ receiving antenna $R{x_j}$ is expressed as
\[d_{ji}^{DA} = \sqrt {d_{TR}^2 + {{\left( {\left| {i - j} \right|{d_T} + {d_{TR}}\tan \gamma \cos \eta  + \delta \cos \theta } \right)}^2} + {{\left( {{r_{\max }}(z) + {d_{TR}}\tan \gamma \sin \eta  + \delta \sin \theta } \right)}^2}}. \tag{18}\]
The channel response between the $i - th$ transmitting antenna $T{x_i}$ and the $i - th$ receiving antenna $R{x_i}$ is given by
\[h_{ii}^l = \beta \frac{\lambda }{{4\pi d_{ii}^{DA}}}{e^{ - ikd_{ii}^{DA}}}{e^{ - i\frac{\pi }{2}l}}. \tag{19}\]
The channel response between the $i - th$ transmitting antenna $T{x_i}$ and the $j - th$ receiving antenna $R{x_j}$ is given by
\[h_{ji}^l = {\beta _{ji}}\frac{\lambda }{{4\pi d_{ji}^{DA}}}{e^{ - ikd_{ji}^{DA}}}{e^{ - i\phi _{ji}^{DA}l}}. \tag{20}\]
Similar to the derivation of section II A, ${\beta _{ji}}$ is expressed as
\[{\beta _{ji}} = \beta \frac{{d_{ji}^{DA}}}{{d_{ii}^{DA}}}{\left( {\frac{{{r_{DA,ji}}}}{{{r_{\max }}\left( z \right)}}} \right)^{\left| l \right|}}{e^{ - \frac{{r_{DA,ji}^2 - r_{\max }^2\left( z \right)}}{{\omega _l^2\left( z \right)}}}}{e^{ - i\frac{{\pi \left( {r_{DA,ji}^2 - r_{\max }^2\left( z \right)} \right)}}{{\lambda {R_l}\left( z \right)}}}}{e^{ik\left( {d_{ji}^{DA} - d_{ii}^{DA}} \right)}}. \tag{21}\]
Furthermore, the channel response between the $i - th$ transmitting antenna $T{x_i}$ and the $j - th$ receiving antenna $R{x_j}$ can be derived as
\[h_{ji}^l = \beta \frac{\lambda }{{4\pi d_{ii}^{DA}}}{\left( {\frac{{{r_{DA,ji}}}}{{{r_{\max }}\left( z \right)}}} \right)^{\left| l \right|}}{e^{ - \frac{{r_{DA,ji}^2 - r_{\max }^2\left( z \right)}}{{\omega _l^2\left( z \right)}}}}{e^{ - i\frac{{\pi \left( {r_{DA,ji}^2 - r_{\max }^2\left( z \right)} \right)}}{{\lambda {R_l}\left( z \right)}}}}{e^{ - ikd_{ii}^{DA}}}{e^{ - i\phi _{ji}^{DA}l}}. \tag{22}\]

\section{Purity model}
\label{sec3}

During the propagation of OAM signals in practical atmosphere environments, OAM signals will be influenced by the atmospheric turbulence. Due to the effect of the atmospheric turbulence, a part of the energy of the OAM signals will be redistributed into the adjacent OAM states, which causes the degradation of the OAM transmission system. Since non-Kolmogorov turbulence models are more suitable for the atmospheric motion of the stratosphere and troposphere than Kolmogorov turbulence models \cite{30Tos08}, the isotropic non-Kolmogorov turbulence model is adopted in this paper. The power spectral density for the refractive index fluctuation of the non-Kolmogorov turbulence is expressed as \cite{31Tang16}
\[{\Phi _n}(\kappa ,\alpha ) = A(\alpha )\tilde C_n^2(\alpha )\frac{{{e^{ - \frac{{{\kappa ^2}}}{{\kappa _m^2}}}}}}{{{{\left( {{\kappa ^2} + \kappa _0^2} \right)}^{\frac{\alpha }{2}}}}}, \tag{23a}\]
\[A(\alpha ) = \frac{1}{{4{\pi ^2}}}\Gamma \left( {\alpha  - 1} \right)\cos \left( {\frac{{\alpha \pi }}{2}} \right), \tag{23b}\]
\[\tilde C_n^2(\alpha ) = \frac{{ - \Gamma \left( \alpha  \right){{\left( {\frac{k}{z}} \right)}^{\frac{\alpha }{2} - \frac{{11}}{6}}}C_n^2}}{{8{\pi ^2}\Gamma \left( {1 - 0.5\alpha } \right){{\left[ {\Gamma \left( {0.5\alpha } \right)} \right]}^2}\sin \left( {0.25\pi \alpha } \right)A\left( \alpha  \right)}}, \tag{23c}\]
where $\alpha $ denotes the generalized spectral index, $3 < \alpha  < 4$, $\Gamma \left(  \cdot  \right)$ denotes the gamma function, $\kappa $ denotes the scalar wave number, $0 \le \kappa  < \infty $. The inter scale parameter is ${\kappa _m} = \frac{{c(\alpha )}}{{{L_i}}} = \frac{1}{{{L_i}}}{\left[ {\frac{{2\pi }}{3}\Gamma \left( {\frac{{5 - \alpha }}{2}} \right)A(\alpha )} \right]^{\frac{1}{{\alpha  - 5}}}}$, where ${L_i}$ denotes the inter scale. The outer scale parameter is ${\kappa _0} = \frac{{2\pi }}{{{L_o}}}$, where ${L_o}$ denotes the outer scale. $\tilde C_n^2(\alpha )$ denotes the generalized refractive index structure parameter of the non-Kolmogorov turbulence. $C_n^2$ denotes the refractive index structure constant of the Kolmogorov turbulence and the turbulence strength increases with the increase of $C_n^2$ \cite{8Sun16}. The proposed OAM-MIMO transmission system is deployed to transmit the OAM signals near the ground at a short propagation distance, so there is little change in temperature and humidity during the transmission path. In addition, the value of $C_n^2$ depends on height, humidity and temperature \cite{32Li16}, so the value of $C_n^2$ remains the same during the transmission path. The value of $C_n^2$ of the millimeter-wave near the ground are ranging from $1.6 \times {10^{ - 12}}{\rm{ }}{{\rm{m}}^{ - \frac{{\rm{2}}}{{\rm{3}}}}}$ to $5.5 \times {10^{ - 12}}{\rm{ }}{{\rm{m}}^{ - \frac{{\rm{2}}}{{\rm{3}}}}}$ and the statistical average value of $C_n^2$ is $3.9 \times {10^{ - 12}}{\rm{ }}{{\rm{m}}^{ - \frac{{\rm{2}}}{{\rm{3}}}}}$ \cite{33Mcmillan08}, \cite{34Hill88}.

The distortion of the wavefront of OAM beams caused by the atmospheric turbulence can lead to the loss of the orthogonality among different OAM states \cite{35Amphawan19}. Based on the Rytov approximation, the phase distortion induced by the atmospheric turbulence is expressed as
\[u(r,\phi ,z) = u_{DA}^{p,l}(r,\phi ,z){e^{\psi (r,\phi ,z)}}, \tag{24a}\]
with
\[\left\langle {{e^{\psi (r,\phi ,z) + {\psi ^*}(r,\phi ',z)}}} \right\rangle  = {e^{ - \frac{1}{2}D(\rho ,z)}} = {e^{ - {{\left( {\frac{\rho }{{{\rho _0}}}} \right)}^{\alpha  - 2}}}}, \tag{24b}\]
\[{\rho _0} = {\left[ {\frac{{D(\rho ,z)}}{{2{\rho ^{\alpha  - 2}}}}} \right]^{\frac{1}{{2 - \alpha }}}}, \tag{24c}\]
where $\left\langle {{e^{\psi (r,\phi ,z) + {\psi ^*}(r,\phi ',z)}}} \right\rangle $ denotes the normalized mutual coherence function, which indicates the loss of spatial coherence caused by the atmospheric turbulence. $\rho $ is the distance between $(r,\phi ,z)$ and $(r,\phi ',z)$, ${\rho ^2} = 2{r^2}[1 - \cos (\phi  - \phi ')]$. ${\rho _0}$ denotes the spatial coherence radius of the Gaussian beam wave. Since the LG beam is a high-order Gaussian beam, the spatial coherence radius of the LG beam can be denoted as ${\rho _0}$. The wave structure function $D(\rho ,z)$ in (24c) is extended as \cite{36Andrews05}
\[D(\rho ,z) = 8{\pi ^2}{k^2}z\int_0^1 {\int_0^\infty  {\kappa {\Phi _n}(\kappa ,\alpha ){e^{ - \frac{{\Lambda z{\kappa ^2}{\xi ^2}}}{k}}}\left\{ {{I_0}(\Lambda \xi \kappa \rho ) - {J_0}[(1 - \bar \Theta \xi )\kappa \rho ]} \right\}d\kappa } d\xi }, \tag{25}\]
where $\bar \Theta  =  - \frac{z}{{{R_l}(z)}}$ denotes the complementary parameter of the LG beam at the receiver. $\Lambda  = \frac{{2z}}{{k{w_l}^2(z)}}$ denotes the diffraction parameter of the LG beam at the receiver. $\xi $ denotes the normalized distance variable, $0 \le \xi  \le 1$. Based on the expansion and asymptotic formula of the Bessel function of order $m$ \cite{37Andrews11}, (25) can be rewritten as
\[D(\rho ,z) \approx  - {\pi ^2}{2^{3 - \alpha }}\alpha {k^2}zA(\alpha )\tilde C_n^2(\alpha )\frac{{{{(1 - \bar \Theta )}^{\alpha  - 1}} - 1}}{{\bar \Theta (\alpha  - 1)}}\frac{{\Gamma \left( { - \frac{\alpha }{2}} \right)}}{{\Gamma \left( {\frac{\alpha }{2}} \right)}}{\rho ^{\alpha  - 2}}. \tag{26}\]
Based on (26), ${\rho _0}$ can be expressed as
\[{\rho _0} \approx 2{\left[ { - {\pi ^2}\alpha {k^2}zA(\alpha )\tilde C_n^2(\alpha )\frac{{{{(1 - \bar \Theta )}^{\alpha  - 1}} - 1}}{{\bar \Theta (\alpha  - 1)}}\frac{{\Gamma \left( { - \frac{\alpha }{2}} \right)}}{{\Gamma \left( {\frac{\alpha }{2}} \right)}}} \right]^{\frac{1}{{2 - \alpha }}}}. \tag{27}\]
Based on the quadratic approximation \cite{38Zhang16}, (24b) satisfies
\[\left\langle {{e^{\psi (r,\phi ,z) + {\psi ^*}(r,{\phi '},z)}}} \right\rangle  \approx {e^{ - {{\left( {\frac{\rho }{{\left| {{\rho _0}} \right|}}} \right)}^2}}} = {e^{\frac{{2{r^2}[\cos (\phi  - \phi ') - 1]}}{{{{\left| {{\rho _0}} \right|}^{\rm{2}}}}}}}. \tag{28}\]
Since the misalignment between the transmitting and receiving antennas will cause a part of the energy of the OAM signals redistributed into the adjacent OAM states. The LG beam with unaligned antennas but no effect of the atmospheric turbulence in (15) is expressed as
\[u_{DA}^{p,l}(r,\phi ,z) = \frac{1}{{\sqrt {2\pi } }}\sum\limits_{l =  - \infty }^\infty  {{a_l}(r,z){e^{ - il\phi }}}, \tag{29a}\]
with
\begin{equation}
\begin{split}
  {a_l}(r,z)= &\frac{1}{{\sqrt {2\pi } }}\int_0^{2\pi } {u_{DA}^{p,l}(r,\phi ,z){e^{il\phi }}} d\phi \hfill \\
= &\int_0^{2\pi } {\frac{a}{\pi }\sqrt {\frac{{p!}}{{2(p + \left| l \right|)!}}} \frac{1}{{\omega _l^2(z)}}{{\left( {\frac{{\sqrt 2 r}}{{{\omega _l}(z)}}} \right)}^{\left| l \right|}}{e^{ - {{\left( {\frac{r}{{{\omega _l}(z)}}} \right)}^2}}}} L_p^{\left| l \right|}\left( {\frac{{2{r^2}}}{{\omega _l^{\text{2}}(z)}}} \right) \hfill \\
  & \times {e^{i(\left| l \right| + 2p + 1)\zeta (z)}}{e^{\frac{{ - i\pi {r^2}}}{{\lambda {R_l}(z)}}}}{\left( {r{e^{i\phi }} - \delta {e^{i\theta }}} \right)^l}\exp \left( { - \frac{{{r^2} + {\delta ^2}}}{{\omega _l^2\left( z \right)}}} \right) \hfill \\
  & \times \sum\limits_{m =  - \infty }^\infty  {{I_m}\left( {\frac{{2r\delta }}{{\omega _l^2\left( z \right)}}} \right)} \sum\limits_{n =  - \infty }^\infty  {{J_n}} \left( {\varsigma r} \right)\exp \left[ {im\left( {\phi  - \theta } \right) + in\left( {\phi  - \eta  + \frac{\pi }{2}} \right)} \right]d\phi \\
\end{split}. \tag{29b}
\end{equation}
When the signals is affected by the atmospheric turbulence, the distorted electric field of (24a) can be expanded as
\[u(r,\phi ,z) = \frac{1}{{\sqrt {2\pi } }}\sum\limits_{l =  - \infty }^\infty  {{b_l}(r,z){e^{ - il\phi }}}. \tag{30}\]
Based on the discrete-time Fourier transform, ${b_l}(r,z)$ is expressed as
\[{b_l}(r,z) = \frac{1}{{\sqrt {2\pi } }}\int_0^{2\pi } {u(r,\phi ,z){e^{il\phi }}} d\phi. \tag{31}\]
When the transmitted OAM state is ${l_j}$, the received distorted OAM state is $l$, then the power weight of signal is expressed as
\[{T_l}({l_j},z) = \frac{{{A_l}({l_j},z)}}{{\sum\limits_{n =  - \infty }^\infty  {{A_n}({l_j},z)} }}, \tag{32}\]
with
\[\begin{aligned}
{A_l}({l_j},z) =& \int_{{r_{\max }}(z)}^{{r_{\max }}(z) + dr} {\left\langle {{{\left| {{b_l}(r,z)} \right|}^2}} \right\rangle rdr} \\
 =& \frac{{{a^2}\omega _{{l_j}}^{ - 4}(z)p!}}{{2{\pi ^2}(p + |{l_j}|)!}}\int_{{r_{\max }}(z)}^{{r_{\max }}(z) + dr} {{e^{2\left( { - \frac{{{r^2} + {\delta ^2}}}{{\omega _{{l_j}}^2\left( z \right)}}} \right)}}{{\left( {\frac{{2{r^2}}}{{\omega _{{l_j}}^2(z)}}} \right)}^{|{l_j}|}}} {e^{ - \frac{{2{r^2}}}{{\omega _{{l_j}}^2(z)}} - \frac{{2{r^2}}}{{{{\left| {{\rho _0}} \right|}^{\rm{2}}}}}}}\\
 &\times {\left[ {L_p^{|{l_j}|}\left( {\frac{{2{r^2}}}{{\omega _{{l_j}}^2(z)}}} \right)} \right]^2}\int_0^{2\pi } {\int_0^{2\pi } {{e^{i(l - {l_j})(\phi  - \phi ')}}} } {e^{\frac{{2{r^2}\cos (\phi  - \phi ')}}{{{{\left| {{\rho _0}} \right|}^{\rm{2}}}}}}}{\left( {r{e^{i\phi }} - \delta {e^{i\theta }}} \right)^{2{l_j}}}\\
 &\times {\left( {\sum\limits_{m =  - \infty }^\infty  {{I_m}\left( {\frac{{2r\delta }}{{\omega _{{l_j}}^2\left( z \right)}}} \right)\sum\limits_{n =  - \infty }^\infty  {{J_n}} \left( {\varsigma r} \right)} {e^{\left[ {im\left( {\phi  - \theta } \right) + in\left( {\phi  - \eta  + \frac{\pi }{2}} \right)} \right]}}} \right)^2}d\phi d\phi 'rdr
\end{aligned}. \tag{33}\]
$\left\langle {{{\left| {{b_l}(r,z)} \right|}^2}} \right\rangle $ denotes the probability density function of the OAM signals in non-Kolmogorov turbulent flow with OAM state $l$. The integration interval $\left[ {{r_{\max }}(z),{r_{\max }}(z) + dr} \right]$ stands for the position range of the $j - th$ receiving antenna $R{x_j}$. Similarly, $\sum\limits_{n =  - \infty }^\infty  {{A_n}({l_j},z)} $ is expressed as
\[\begin{aligned}
\sum\limits_{n =  - \infty }^\infty  {{A_n}({l_j},z)}  =& \int_{{r_{\max }}(z)}^{{r_{\max }}(z) + dr} {\left\langle {{{\left| {{a_{{l_j}}}(r,z)} \right|}^2}} \right\rangle rdr} \\
 =& \int_{{r_{\max }}(z)}^{{r_{\max }}(z) + dr} {\int_0^{2\pi } {\frac{{{a^2}p!}}{{2\pi {}^2(p + \left| {{l_j}} \right|)!}}\frac{1}{{\omega _{{l_j}}^4(z)}}} } {\left( {\frac{{{\rm{2}}{r^{\rm{2}}}}}{{\omega _{{l_j}}^2(z)}}} \right)^{\left| {{l_j}} \right|}}{e^{ - \frac{{{\rm{2}}{r^2}}}{{\omega _{{l_j}}^2(z)}}}}\\
 &\times {\left[ {\sum\limits_{m =  - \infty }^\infty  {{I_m}\left( {\frac{{2r\delta }}{{\omega _{{l_j}}^2\left( z \right)}}} \right)} \sum\limits_{n =  - \infty }^\infty  {{J_n}} \left( {\varsigma r} \right)\exp \left[ {im\left( {\phi  - \theta } \right) + in\left( {\phi  - \eta  + \frac{\pi }{2}} \right)} \right]} \right]^2}\\
 &\times {\left[ {L_p^{|{l_j}|}\left( {\frac{{2{r^2}}}{{\omega _{{l_j}}^2(z)}}} \right)} \right]^2}{\left( {r{e^{i\phi }} - \delta {e^{i\theta }}} \right)^{2{l_j}}}{e^{2\left( { - \frac{{{r^2} + {\delta ^2}}}{{\omega _{{l_j}}^2\left( z \right)}}} \right)}}d\phi rdr{\rm{ }}
\end{aligned}. \tag{34}\]
Based on $\int_0^{{\rm{2}}\pi } {{e^{ - in\phi ' + \eta \cos (\phi ' - \phi )}}d\phi ' = 2\pi {e^{ - in\phi }}{I_m}(\eta )} $ \cite{39Gradshteyn00}, the power weight is derived as
\[{T_l}({l_j},z) = \frac{{{A_l}({l_j},z)}}{{\sum\limits_{n =  - \infty }^\infty  {{A_n}({l_j},z)} }} = \frac{{4{\pi ^2}\int_{{r_{\max }}(z)}^{{r_{\max }}(z) + dr} {F\left( r \right){e^{ - \frac{{2{r^2}}}{{{{\left| {{\rho _0}} \right|}^{\rm{2}}}}}}}{I_{l - {l_j}}}\left( {\frac{{2{r^2}}}{{{{\left| {{\rho _0}} \right|}^2}}}} \right)rdr} }}{{\int_{{r_{\max }}(z)}^{{r_{\max }}(z) + dr} {F\left( r \right)rdr} }}, \tag{35a}\]
\[\begin{aligned}
F\left( r \right) =& \int_0^{2\pi } {{{\left( {\frac{{{\rm{2}}{r^{\rm{2}}}}}{{\omega _{{l_j}}^2(z)}}} \right)}^{\left| {{l_j}} \right|}}} {e^{ - \frac{{{\rm{2}}{r^2}}}{{\omega _{{l_j}}^2(z)}}}}{\left[ {L_p^{|{l_j}|}\left( {\frac{{2{r^2}}}{{\omega _{{l_j}}^2(z)}}} \right)} \right]^2} {\left( {r{e^{i\phi }} - \delta {e^{i\theta }}} \right)^{2{l_j}}}{e^{2\left( { - \frac{{{r^2} + {\delta ^2}}}{{\omega _{{l_j}}^2\left( z \right)}}} \right)}}\\
 &\times {\left[ {\sum\limits_{m =  - \infty }^\infty  {{I_m}\left( {\frac{{2r\delta }}{{\omega _{{l_j}}^2\left( z \right)}}} \right)}  \sum\limits_{n =  - \infty }^\infty  {{J_n}} \left( {\varsigma r} \right){e^{\left[ {im\left( {\phi  - \theta } \right) + in\left( {\phi  - \eta  + \frac{\pi }{2}} \right)} \right]}}} \right]^2}d\phi
\end{aligned}. \tag{35b}\]

\section{Capacity model of OAM-MIMO transmission systems with unaligned antennas}
\label{sec4}

\subsection{Capacity and error probability model}

Assuming that the $j - th$ transmitting antenna $T{x_j}$ transmits the OAM signal ${x_{j,l}}$ with OAM state $l$. At the $j - th$ receiving antenna $R{x_j}$, the received signal is expressed as
\[{y_{j,l}} = \sqrt {{\rho _{jj,ll}}} h_{jj}^l{x_{j,l}} + {I_{j,l}} + {\omega _j}, \tag{36a}\]
\[{I_{j,l}} = \sum\limits_{l' \ne l} {\sqrt {{\rho _{jj,ll'}}} h_{jj}^l{x_{j,l'}}}  + \sum\limits_{j' \ne j,j' = 1}^N {\sqrt {{\rho _{jj',ll}}} h_{jj'}^l{x_{j',l}}}, \tag{36b}\]
where ${\rho _{jj,ll}}$ is the power of the signal received by the receiving antenna $R{x_j}$ with OAM state $l$ and transmitted by the transmitting antenna $T{x_j}$ with OAM state $l$. ${\rho _{jj,ll'}}$ is the power of the distorted signal received by the receiving antenna $R{x_j}$ with OAM state $l$ and transmitted by the transmitting antenna $T{x_j}$ with OAM state $l'$. ${\rho _{jj',ll}}$ is the power of the signal received by the receiving antenna $R{x_j}$ with OAM state $l$ and transmitted by the transmitting antenna $T{x_{j'}}$ with OAM state $l$. ${I_{j,l}}$ denotes the interference signals, ${\omega _j}$ denotes the additional Gaussian white noise with mean zero and variance $\sigma _\omega ^2$.

The interference from other antennas and OAM states causes the degradation of the OAM-MIMO transmission system, the parallel interference cancellation method is used to alleviate the interference of the OAM-MIMO transmission system. Assuming that the received signals ${{\bf{\hat x}}_l} = {\left[ {{{\hat x}_{l,1}}, \cdots ,{{\hat x}_{l,N}}} \right]^T}$ after equalization are expressed as
\[{{\bf{\hat x}}_l} = {{\bf{W}}_l}{{\bf{H}}_l}{{\bf{x}}_l} + {{\bf{W}}_l}{{\bf{I}}_l} + {{\bf{W}}_l}{\bf{\omega }}, \tag{37a}\]
with
\[{{\bf{H}}_l} = \left[ {\begin{array}{*{20}{c}}
{\sqrt {{\rho _{11,ll}}} h_{11}^l}& \cdots &{\sqrt {{\rho _{1N,ll}}} h_{1N}^l}\\
 \vdots &{\sqrt {{\rho _{jj,ll}}} h_{jj}^l}& \vdots \\
{\sqrt {{\rho _{N1,ll}}} h_{N1}^l}& \cdots &{\sqrt {{\rho _{NN,ll}}} h_{NN}^l}
\end{array}} \right], \tag{37b}\]
\[{{\bf{x}}_l} = {\left[ {{x_{1,l}}, \cdots ,{x_{N,l}}} \right]^T}, \tag{37c}\]
\[{{\bf{I}}_l} = {\left[ {{I_{1,l}}, \cdots ,{I_{N,l}}} \right]^T}, \tag{37d}\]
\[{\bf{\omega }} = {\left[ {{\omega _1}, \cdots ,{\omega _N}} \right]^T}, \tag{37e}\]
where ${{\bf{W}}_l}$ denotes the equalization matrix, when the minimum mean square error (MMSE) method is used here. ${{\bf{W}}_l}$ is expressed as ${{\bf{W}}_l} = {\left( {{\bf{H}}_l^{\rm{H}}{{\bf{H}}_l} + \frac{{{G_n}}}{{{G_x}}}{{\rm{I}}_N}} \right)^{ - 1}}{\bf{H}}_l^{\rm{H}}$, where ${G_x}$, ${G_n}$ and ${{\rm{I}}_N}$ stand for the transmit power, noise power and identity matrix, respectively. Then the signal-to-interference-and-noise ratio (SINR) of the receiving antenna $R{x_j}$ is given by
\[\gamma _j^l = \frac{{{G_{D,j,l}}}}{{{G_{I,j,l}} + {G_{N,j,l}}}}, \tag{38a}\]
with \[{G_{D,j,l}} = {\rho _{jj,ll}}{\left| {{w_{j,l}}h_{jj}^l} \right|^2}, \tag{38b}\]
\[{G_{I,j,l}} = \sum\limits_{j' \ne j,j' = 1}^N {{\rho _{jj',ll}}{{\left| {{w_{j,l}}h_{jj'}^l} \right|}^2}}  + \sum\limits_{l' \ne l} {{\rho _{jj,ll'}}{{\left| {{w_{j,l}}h_{jj}^l} \right|}^2}} , \tag{38c}\]
\[{G_{N,j,l}} = {\left| {{w_{j,l}}} \right|^2}\sigma _\omega ^2. \tag{38d}\]

To obtain the error probability of the OAM-MIMO transmission system, the probability of the correct demodulation of the signals need to be derived first. To demodulate the signals correctly, the OAM states of the signals need to be estimated correctly, then the signals transmitted by the transmitting antennas should be estimated correctly. The correct demodulation of the OAM states of the transmitted signals is denoted as ${\Delta _{{\rm{OAM}}}}$ and the correct demodulation of the transmitted signals is denoted as ${\Delta _{{\rm{sig}}}}$. When the total correct demodulation is denoted as $\Delta $, then the probability of $\Delta $ is expressed as
\[P\left( \Delta  \right) = P\left( {{\Delta _{{\rm{sig}}}}\left| {{\Delta _{{\rm{OAM}}}}} \right.} \right)P\left( {{\Delta _{{\rm{OAM}}}}} \right), \tag{39}\]
where $P\left(  \cdot  \right)$ denotes the probability of an event, $P\left( {{\Delta _{{\rm{sig}}}}\left| {{\Delta _{{\rm{OAM}}}}} \right.} \right)$ is the conditional correct estimation probability of the transmitted signals while the OAM states have already been demodulated correctly. $P\left( {{\Delta _{{\rm{OAM}}}}} \right)$ denotes the correct estimation probability of the OAM states of the transmitted signals.

The pairwise error probability (PEP) can be used to calculate $P\left( {{\Delta _{{\rm{OAM}}}}} \right)$. The PEP $P\left( {l \to l'} \right)$ represents the error probability of the signals transmitted with OAM state $l$ but estimated at the receiver with OAM state $l'$. $P\left( {l \to l'} \right)$ can be expressed as \cite{40Jiang20}, \cite{41Bas18}
\[P\left( {l \to l'} \right) = \mathbb{Q}\left( {\sqrt {\frac{{L \cdot {\rho _{jj,ll}}(1 - D)}}{{2\sigma _\omega ^2}}} } \right), \tag{40}\]
where $\mathbb{Q}\left( x \right) = \int\limits_x^\infty  {\frac{1}{{\sqrt {2\pi } }}\exp \left( { - \frac{{{t^2}}}{2}} \right)dt} $, $D = \operatorname{Re} \left\{ {\exp \left[ {i\left( {l - l'} \right)\phi } \right]} \right\}$.
Then $P\left( {{\Delta _{{\text{OAM}}}}} \right)$ is derived as
\[P\left( {{\Delta _{{\text{OAM}}}}} \right) = 1 - P\left( {l \to l'} \right). \tag{41}\]

After the estimation of the OAM states, the maximum likelihood detection is used to demodulate the transmitted signals. Assuming the Q-point constellation modulation is used here, the error probability of the transmitted signals is derived as \cite{42Irshid91}, \cite{43Simon05}
\[{e_{{\text{sig}}}} = \frac{1}{{Q{{\log }_2}Q}}\sum\limits_{q = 1}^Q {\sum\limits_{{q_1} = 1}^Q {\mathbb{D}({x_q},{{\hat x}_{{q_1}}})} \mathbb{Q}\left( {\sqrt {\frac{\rho _{jj,ll}}{{2\sigma _w^2}}{{\left| {{w_{j,l}}h_{ii}^l} \right|}^2}{{\left| {{x_q},{{\hat x}_{{q_1}}}} \right|}^2}} } \right)}, \tag{42}\]
where ${x_q}$ is the $q - th$ input transmitted symbol, ${\hat x_{{q_1}}}$ is the ${q_1} - th$ input transmitted symbol. $\mathbb{D}({x_q},{\hat x_{{q_1}}})$ denotes the Hamming distance between ${x_q}$ and ${\hat x_{{q_1}}}$. The codeword difference between ${x_q}$ and ${\hat x_{{q_1}}}$ is denoted as $\left| {{x_q},{{\hat x}_{{q_1}}}} \right| = 2\mathbb{R}\left[ {{x_q}{{\left( {{x_q} - {{\hat x}_{{q_1}}}} \right)}^ * }} \right]$. Therefore, the correct estimation probability of the transmitted signals is expressed as
\[\mathcal{P}\left( {\left. {{\Delta _{{\text{sig}}}}} \right|{\Delta _{{\text{OAM}}}}} \right) = 1 - {e_{{\text{sig}}}}. \tag{43}\]
Combine (41), (43) with (39), the error probability of the OAM-MIMO transmission system is derived as
\[\begin{aligned}
  P\left( \varepsilon  \right) = & 1 - P\left( \Delta  \right) = 1 - \left( {1 - {e_{{\text{sig}}}}} \right)\left( {1 - P\left( {l \to l'} \right)} \right) \hfill \\
  = & 1 - \left( {1 - \frac{1}{{Q{{\log }_2}Q}}\sum\limits_{q = 1}^Q {\sum\limits_{{q_1} = 1}^Q {\mathbb{D}({x_q},{{\hat x}_{{q_1}}})} \mathbb{Q}\left( {\sqrt {\frac{\rho _{jj,ll}}{{2\sigma _w^2}}{{\left| {{w_{j,l}}h_{ii}^l} \right|}^2}{{\left| {{x_q},{{\hat x}_{{q_1}}}} \right|}^2}} } \right)} } \right)  \hfill \\
  &\times \left( {1 - \mathbb{Q}\left( {\sqrt {\frac{{L \cdot {\rho _{jj,ll}}(1 - D)}}{{2\sigma _\omega ^2}}} } \right)} \right) \hfill \\
\end{aligned}. \tag{44}\]

For the capacity model of the OAM-MIMO transmission system, the capacity of the receiving antenna $R{x_j}$ is expressed as
\[{C_j} = \sum\limits_l {{{\log }_2}\left( {1 + \gamma _j^l} \right)}. \tag{45}\]
Then the total capacity of the OAM-MIMO channel is expressed as
\[\begin{aligned}
  {C_t} &= \sum\limits_j {{C_j}}  = \sum\limits_j {\sum\limits_l {{{\log }_2}\left( {1 + \gamma _j^l} \right)} }  \hfill \\
  & =\sum\limits_j {\sum\limits_l {{{\log }_2}\left( {1 + \frac{{{\rho _{jj,ll}}{{\left| {{w_{j,l}}h_{jj}^l} \right|}^2}}}{{\sum\limits_{j' \ne j,j' = 1}^N {{\rho _{jj',ll}}{{\left| {{w_{j,l}}h_{jj'}^l} \right|}^2}}  + \sum\limits_{l' \ne l} {{\rho _{jj,ll'}}{{\left| {{w_{j,l}}h_{jj}^l} \right|}^2}}  + {{\left| {{w_{j,l}}} \right|}^2}\sigma _\omega ^2}}} \right)} }  \hfill \\
\end{aligned}. \tag{46}\]

\subsection{OAM state interval optimal method}

\begin{algorithm}
\caption{\textbf{Optimization of OAM state interval algorithm}.}
\label{alg1}
\begin{algorithmic}
\State \textbf{Begin:} \begin{enumerate} 
     \item Assuming $o = 0$ to be the initial OAM state interval, $o \leqslant 10$ and $\hat C = 0$;

     \item $o = o + 1$, then determine the OAM states set $S$ according to the value of $L$;

     \item Calculate ${A_l}\left( {{l_j},z} \right)$ and $\sum\limits_{n =  - \infty }^\infty  {{A_n}({l_j},z)} $;

     \item Calculate ${T_l}({l_j},z) = \frac{{{A_l}({l_j},z)}}{{\sum\limits_{n =  - \infty }^\infty  {{A_n}({l_j},z)} }}$;

     \item Calculate $\gamma _j^l = \frac{{{G_{D,j,l}}}}{{{G_{I,j,l}} + {G_{N,j,l}}}}$;

     \item Calculate ${C_t} = \sum\limits_j {\sum\limits_l {{{\log }_2}\left( {1 + \gamma _j^l} \right)} } $, if ${C_t} \geqslant \hat C$, then $\hat C = {C_t}$, otherwise $\hat C$ remains the same;

     \item Return to step 2 and repeat step 2 to step 6;

     \item Find the optimal OAM state interval $\hat o$ and the corresponding ${C_t}$ according to the iterations above.

                       \end{enumerate}
\State \textbf{end Begin}
\end{algorithmic}
\end{algorithm}

Based on (35a), ${T_l}({l_j},z)$ changes with ${I_{l - {l_j}}}\left(  \cdot  \right)$ when $l - {l_j}$ changes. Based on the character of the Bessel function of the first kind, ${I_{l - {l_j}}}\left(  \cdot  \right)$ decreases with the increase of $l - {l_j}$ which leads to the decrease of ${T_l}({l_j},z)$. Then ${\rho _{jj,ll'}}$ decreases and $\gamma _j^l$ increases. In the end, the system capacity increases with the decrease of ${\rho _{jj,ll'}}$. Moreover, the increase of $l - {l_j}$ means a larger OAM state to be transmitted which results in the decrease of ${\rho _{jj,ll}}$, the decrease of ${\rho _{jj,ll}}$ leads to the decrease of $\gamma _j^l$ which decreases the system capacity. Therefore, there exists an optimized value of $l - {l_j}$, {\em i.e.}, OAM state interval to maximize the system capacity. Define the optimized OAM state interval as $\hat o$ , then the optimization equation is given by
\[\hat o = \arg \mathop {\max }\limits_{o > 0} {C_t}. \tag{47}\]

Considering the practical OAM-MIMO transmission system, the OAM state interval is limited to $o \leqslant 10$. The Algorithm 1 is developed to improve the capacity of OAM-MIMO transmission systems with unaligned antennas.

\section{Numerical results}
\label{sec5}

\begin{figure}
  \centering
  \includegraphics[width=13cm]{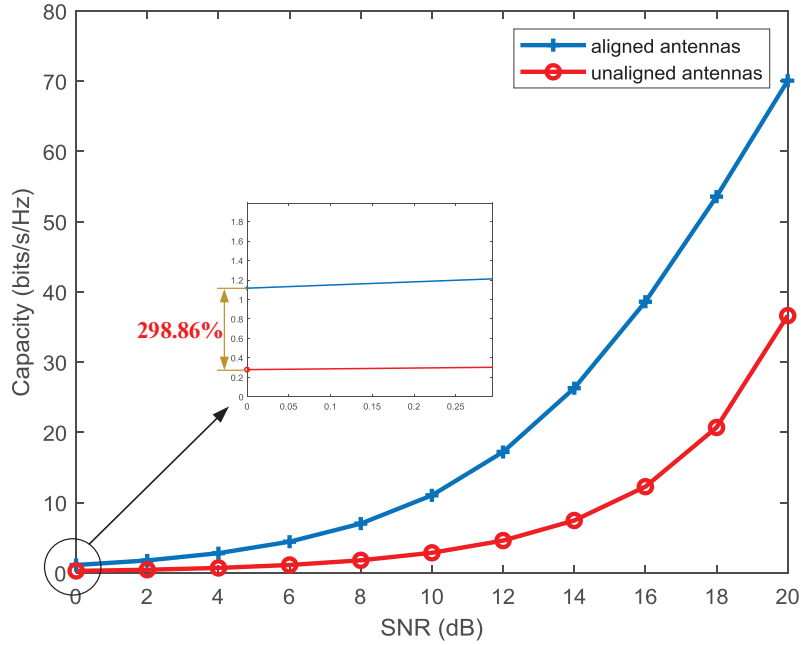}
  \caption{Capacity of the OAM-MIMO transmission system with respect to SNR with aligned and unaligned antennas.}\label{fig3}
\end{figure}

In this section, the capacity and error probability of the proposed OAM-MIMO transmission system are simulated for performance analysis. The default simulation parameters of OAM-MIMO transmission systems are configured as follows: the wavelength is $\lambda  = 0.005\,\operatorname{m} $, the number of OAM states simultaneously transmitted by one antenna is $L = 4$, the propagation distance $z$ is $50\,\operatorname{m} $, the number of antennas is $N = 8$, the refractive index structure constant is $C_n^2 = 3 \times {10^{ - 12}}\,{\operatorname{m} ^{ - \frac{2}{3}}}$, the generalized spectral index is $\alpha  = 3.7$, the radial displacement is $\delta  = \lambda $, the azimuth angle of the radial displacement is $\theta  = \frac{\pi }{2}$, the angle of deflection is $\gamma  = {10^{ - 4}}$, the azimuth angle of the deflection is $\eta  = 0$ \cite{17Vas05}, the signal-to-noise ratio is $SNR = 10\,\operatorname{dB} $.

\begin{figure}
  \centering
  \includegraphics[width=16cm]{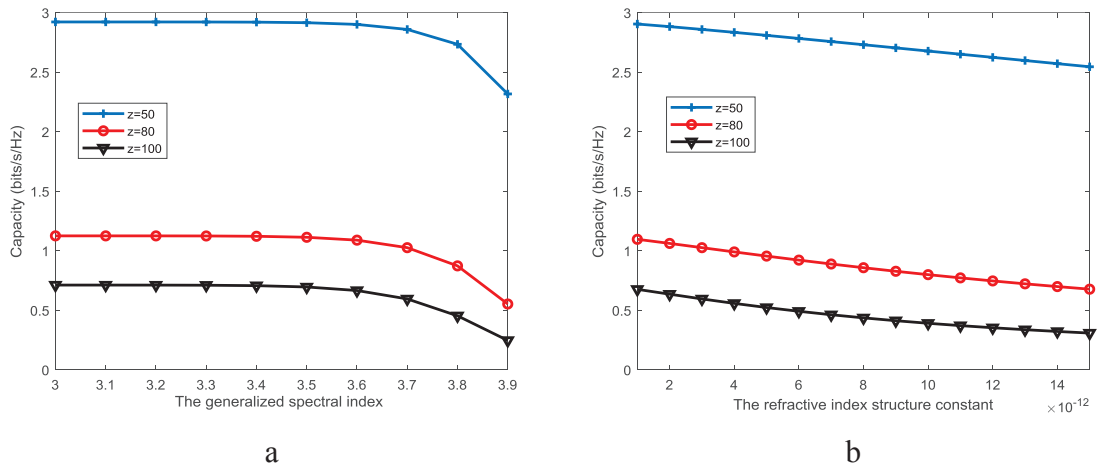}
  \caption{Capacity of the OAM-MIMO transmission system with respect to the generalized spectral index and the refractive index structure constant considering different propagation distances.}\label{fig4}
\end{figure}

\begin{figure}
  \centering
  \includegraphics[width=13cm]{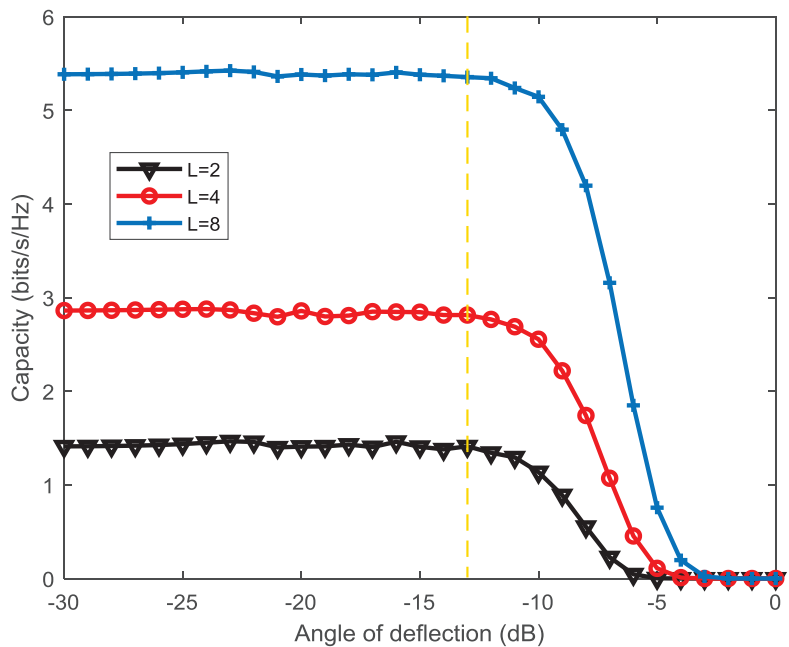}
  \caption{Capacity of the OAM-MIMO transmission system with respect to the angle of deflection considering different number of OAM states.}\label{fig5}
\end{figure}

In Fig. 3, the capacity of the OAM-MIMO transmission system with respect to SNR with aligned and unaligned antennas is analyzed. As shown in Fig. 3, the capacities of the OAM-MIMO transmission system with aligned and unaligned antennas both increase with the increase of SNR. The capacity of the OAM-MIMO transmission system with aligned antennas is larger than the capacity of the OAM-MIMO transmission system with unaligned antennas. When the SNR is 0 dB, compared to the OAM-MIMO transmission system with unaligned antennas, the OAM-MIMO transmission system with aligned antennas achieves up to 298.86\% improvements of the capacity. The results of Fig. 3 indicate that the misalignment between antennas decreases the capacity of the OAM-MIMO transmission system.

\begin{figure}
  \centering
  \includegraphics[width=13cm]{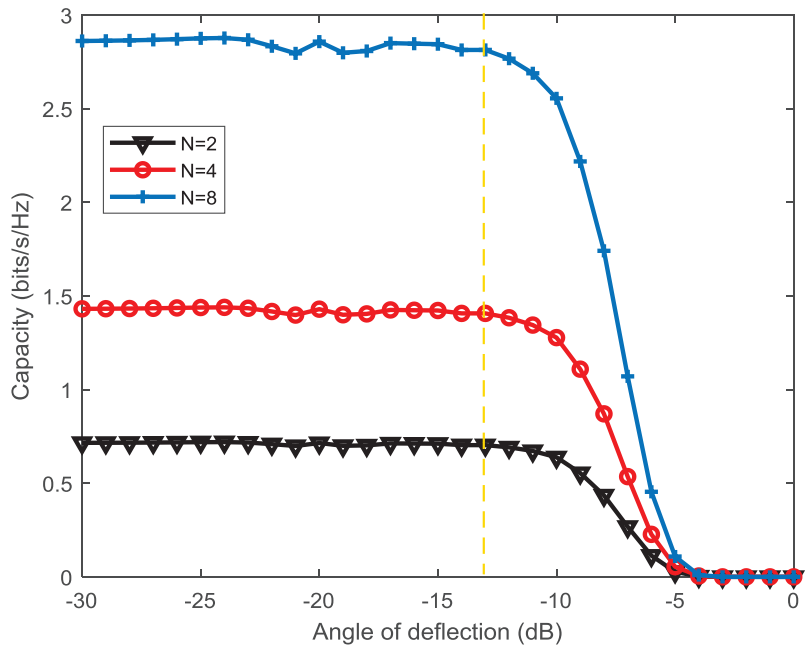}
  \caption{Capacity of the OAM-MIMO transmission system with respect to the angle of deflection considering different number of antennas.}\label{fig6}
\end{figure}

In Fig. 4(a), the capacity of the OAM-MIMO transmission system with respect to the generalized spectral index considering different propagation distances is analyzed. Based on the weak fluctuation condition, the change range of the generalized spectral index is configured as $\left[ {3,3.9} \right]$. As shown in Fig. 4(a), the capacity of the OAM-MIMO transmission system decreases with the increase of the generalized spectral index. When the generalized spectral index is fixed, the capacity of the OAM-MIMO transmission system increases with the decrease of the propagation distance. In Fig. 4(b), the capacity of the OAM-MIMO transmission system with respect to the refractive index structure constant considering different propagation distances is analyzed. As shown in Fig. 4(b), the capacity of the OAM-MIMO transmission system decreases with the increase of the refractive index structure constant.

\begin{figure}
  \centering
  \includegraphics[width=13cm]{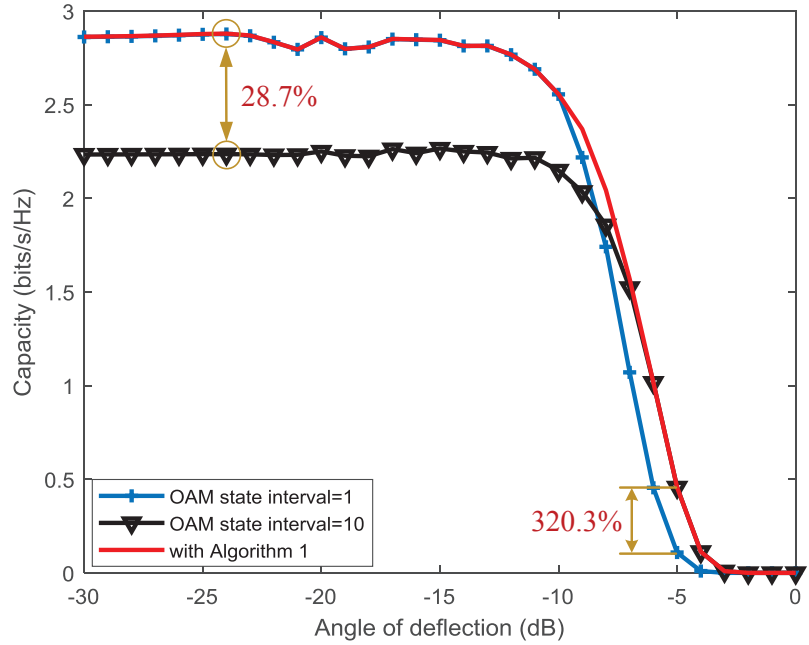}
  \caption{Capacity of the OAM-MIMO transmission system with respect to the angle of deflection considering the optimal OAM state interval.}\label{fig7}
\end{figure}

In Fig. 5, the capacity of the OAM-MIMO transmission system with respect to the angle of deflection considering different number of OAM states is analyzed. As shown in Fig. 5, when the angle of deflection is less than or equal to -13 dB (shown as the yellow dotted line), the capacity of the OAM-MIMO transmission system approaches a saturation value. When the angle of deflection is larger than -13 dB, the capacity of the OAM-MIMO transmission system decreases with the increase of the angle of deflection. When the angle of deflection is fixed, the capacity of the OAM-MIMO transmission system increases with the increase of the number of OAM states.

In Fig. 6, the capacity of the OAM-MIMO transmission system with respect to the angle of deflection considering different number of antennas is analyzed. As shown in Fig. 6, when the angle of deflection is fixed, the capacity of the OAM-MIMO transmission system increases with the increase of the number of antennas.

\begin{figure}
  \centering
  \includegraphics[width=13cm]{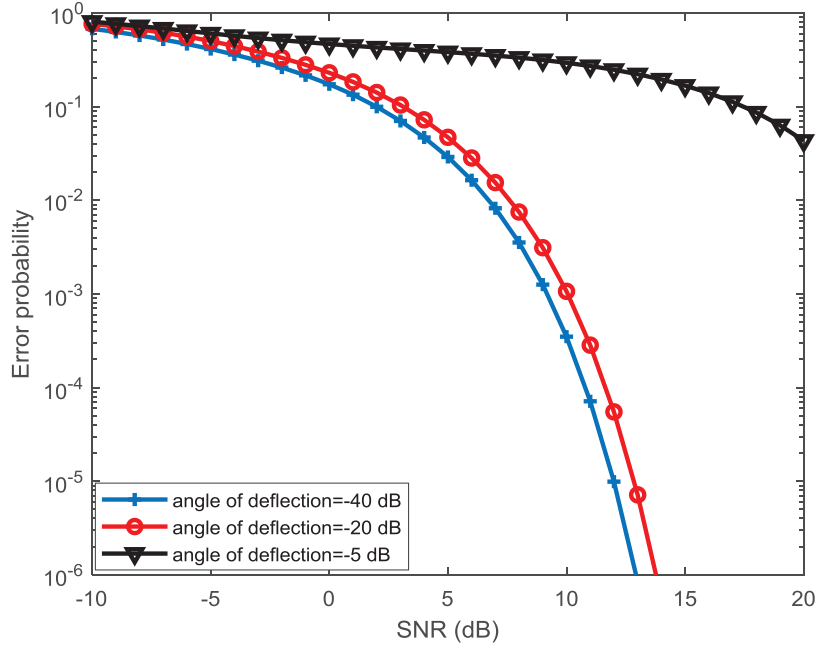}
  \caption{Error probability of the OAM-MIMO transmission system with respect to SNR considering different angles of deflection.}\label{fig8}
\end{figure}

In Fig. 7, the capacity of the OAM-MIMO transmission system with respect to the angle of deflection considering the optimal OAM state interval is analyzed. As shown in Fig. 7, when the angle of deflection is less than -8 dB, the capacity of the OAM-MIMO transmission system decreases with the increase of the OAM state interval. When the angle of deflection is larger than or equal to -8 dB, the capacity of the OAM-MIMO transmission system increases with the increase of the OAM state interval. The capacity of the OAM-MIMO transmission system with the proposed optimization algorithm outperforms the capacity of the OAM-MIMO transmission system without the proposed optimization algorithm. Especially, when the angle of deflection is -24 dB, the optimal OAM state interval $\hat o$ improves the capacity of the OAM-MIMO transmission system by up to 28.7\% compared to the OAM state interval $o=10$. When the angle of deflection is -5 dB, the optimal OAM state interval $\hat o$ improves the capacity of the OAM-MIMO transmission system by up to 320.3\% compared to the OAM state interval $o=1$.

\begin{figure}
  \centering
  \includegraphics[width=13cm]{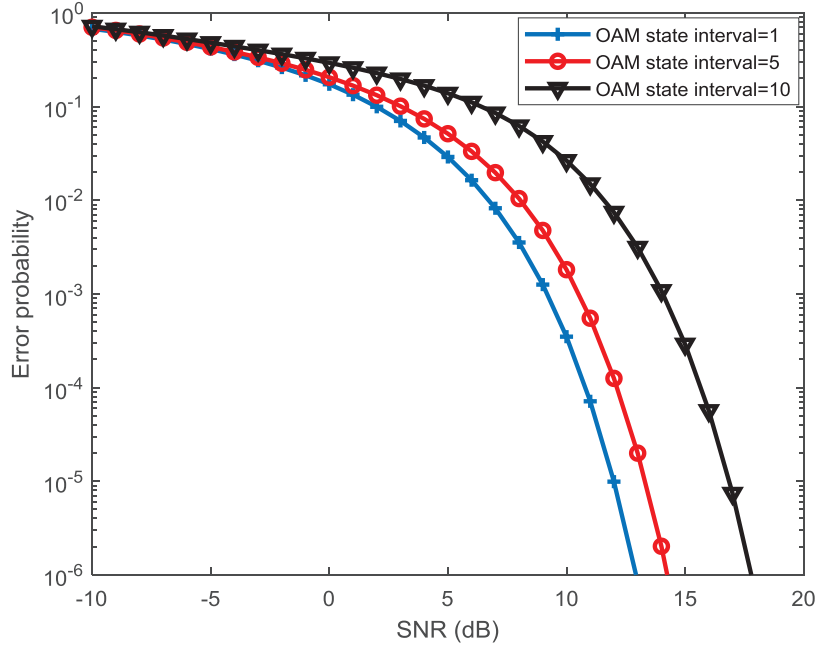}
  \caption{Error probability of the OAM-MIMO transmission system with respect to SNR considering different OAM state intervals.}\label{fig9}
\end{figure}

In Fig. 8, the error probability of the OAM-MIMO transmission system with respect to SNR considering different angles of deflection is analyzed. As shown in Fig. 8, the error probability of the OAM-MIMO transmission system decreases with the increase of SNR. When the SNR is fixed, the error probability of the OAM-MIMO transmission system increases with the increase of angle of deflection.

In Fig. 9, the error probability of the OAM-MIMO transmission system with respect to SNR considering different OAM state intervals is analyzed. When the SNR is fixed, the error probability of the OAM-MIMO transmission system increases with the increase of the OAM state interval.

\section{Conclusion}
\label{sec6}

In this paper, the influences of the atmospheric turbulence and misalignment between the transmitting and receiving antennas on the OAM-MIMO transmission system is investigated. A new channel model of the OAM-MIMO transmission system with unaligned antennas is proposed. Considering the impacts of unaligned antennas and non-Kolmogorov turbulence, a purity model of the OAM-MIMO transmission system is derived. Moreover, the error probability and capacity models of the OAM-MIMO transmission system are derived. Furthermore, the capacity of the OAM-MIMO transmission system is improved by the proposed optimization algorithm. Numerical results show that the OAM-MIMO transmission system with the optimal OAM state interval $\hat o$ achieves up to 28.7\% and 320.3\% capacity gain when the angle of deflection between the transmitting and receiving antennas is -24 dB and -5 dB, respectively. For future works, the analytical solutions of the optimal OAM state interval are still needed to be investigated for OAM-MIMO transmission systems with unaligned antennas.





\vspace{-6 mm}

\end{document}